\documentclass[structabstract]{aa}  
\usepackage{graphicx}
\usepackage{txfonts}
\usepackage{natbib}
\begin{document}
\title{Fragmentation and disk formation in high-mass star formation:\\ The ALMA view of G351.77-0.54 at $0.06''$ resolution}

%   \subtitle{I. Overviewing the $\kappa$-mechanism}

   \author{H.~Beuther
          \inst{1}
          \and
          A.J.~Walsh
          \inst{2}
          \and
          K.G.~Johnston
          \inst{3}
          \and
          Th.~Henning
          \inst{1}
          \and
          R.~Kuiper
          \inst{4, 1}
          \and
          S.N.~Longmore
          \inst{5}
          \and
          C.M.~Walmsley
          \inst{6}
}
   \institute{$^1$ Max Planck Institute for Astronomy, K\"onigstuhl 17,
              69117 Heidelberg, Germany, \email{beuther@mpia.de}\\
              $^2$ International Centre for Radio Astronomy Research, Curtin University, GPO Box U1987, Perth WA 6845, Australia\\
              $^3$ School of Physics and Astronomy, University of Leeds, Leeds, LS2 9JT, UK\\
              $^4$ University of T\"ubingen, Institute of Astronomy and Astrophysics, Auf der Morgenstelle 10, D-72076 T\"ubingen, Germany\\
              $^5$ Astrophysics Research Institute, Liverpool John Moores University, 146 Brownlow Hill, Liverpool L3 5RF, UK\\
              $^6$ Dublin Institute of Advanced Studies, Fitzwilliam Place 31, Dublin 2, Ireland
}

   \date{Version of \today}

%   \abstract{}
% \abstract{}{}{}{}{} 
% 5 {} token are mandatory
\abstract
  % context heading (optional)
  % {} leave it empty if necessary  
{The fragmentation of high-mass gas clumps and the formation of the
  accompanying accretion disks lie at the heart of high-mass star
  formation research.}
  % aims heading (mandatory)
{We resolve the small-scale structure around the high-mass hot core
  region G351.77-0.54 to investigate its disk and fragmentation
  properties.}
  % methods heading (mandatory)
{Using the Atacama Large Millimeter Array at 690\,GHz with baselines
  exceeding 1.5\,km, we study the dense gas, dust and outflow emission
  at an unprecedented spatial resolution of $0.06''$ (130\,AU
  @2.2\,kpc).}
% results heading (mandatory)
{Within the inner few 1000\,AU, G351.77 fragments into at least four
  cores (brightness temperatures between 58 and 197\,K). The central
  structure around the main submm source \#1 with a diameter of
  $\sim$0.5$''$ does not show additional fragmentation. While the
  CO(6-5) line wing emission shows an outflow lobe in the
  north-western direction emanating from source \#1, the dense gas
  tracer CH$_3$CN shows a velocity gradient perpendicular to the
  outflow that is indicative of rotational motions. Absorption profile
  measurements against the submm source \#2 indicate infall
  rates on the order of $10^{-4}$ to $10^{-3}$\,M$_{\odot}$\,yr$^{-1}$
  which can be considered as an upper limit of the mean accretion
  rates.  The position-velocity diagrams are consistent with a central
  rotating disk-like structure embedded in an infalling envelope, but
  they may also be influenced by the outflow. Using the
      CH$_3$CN($37_k-36_k$) $k-$ladder with excitation temperatures
  up to 1300\,K, we derive a gas temperature map of source \#1
  exhibiting temperatures often in excess of 1000\,K. Brightness
  temperatures of the submm continuum never exceed
  200\,K. This discrepancy between gas temperatures and
  submm dust brightness temperatures (in the optically
  thick limit) indicates that the dust may trace the disk mid-plane
  whereas the gas could be tracing a hotter gaseous disk surface
  layer. In addition, we conduct a pixel-by-pixel Toomre gravitational
  stability analysis of the central rotating structure. The derived
  high $Q$ values throughout the structure confirm that this central
  region appears stable against gravitational instability.}
  % conclusions heading (optional), leave it empty if necessary 
{Resolving for the first time a high-mass hot core at $0.06''$
  resolution at submm wavelengths in the
  dense gas and dust emission allowed us to trace the fragmenting core
  and the gravitationally stable inner rotating disk-like
  structure. A temperature analysis reveals hot gas and
  comparably colder dust that may be attributed to
  different disk locations traced by dust emission and gas lines. The
  kinematics of the central structure \#1 reveal contributions from a
  rotating disk, an infalling envelope and potentially also an
  outflow, whereas the spectral profile toward source \#2 can be
  attributed to infall.}  

\titlerunning{The high-mass core G351.77-0.54 resolved by ALMA at $0.06''$ resolution}

\maketitle

\section{Introduction}
\label{intro}

Understanding the fragmentation of high-mass gas clumps and the
subsequent formation and evolution of accretion disks around young
high-mass protostars is one of the main unsolved questions in
high-mass star formation research from an observational (e.g.,
\citealt{cesaroni2006,sandell2010,keto2010,beltran2011,beuther2006b,beuther2009c,beuther2013b,johnston2015,boley2013,boley2016})
as well as theoretical point of view (e.g.,
\citealt{yorke2002,mckee2003,bonnell2006,krumholz2007b,krumholz2009,smith2009b,kuiper2011,kuiper2013,tan2014}).

Much indirect evidence has been accumulated that suggests high-mass
accretion disks do exist. The main observational argument stems from
collimated molecular outflows, resembling low-mass star formation
(e.g., \citealt{beuther2002d,arce2006,lopez2009}). Such structures are
best explained with underlying high-mass accretion disks driving the
outflows via magneto-centrifugal acceleration.
(Magneto)-hydrodynamical simulations of collapsing high-mass gas cores
in 3D also result in the formation of high-mass accretion disks
\citep{krumholz2006b,kuiper2011,seifried2011,klassen2016}.  There is a
growing consensus in the high-mass star formation community that
accretion disks should also exist, however, it is still unknown
whether such disks are similar to their low-mass counterparts, hence
dominated by the central young stellar object and in Keplerian
rotation, or whether they are self-gravitating non-Keplerian
entities. They may also be a hybrid of both possibilities with a
stable inner Keplerian disk embedded in a larger self-gravitating
non-Keplerian rotating envelope. The existence of disks may be even more
important around high-mass protostars because disks enable the
accretion flow to be shielded against the strong radiation pressure in
the direction of the disks, whereas the radiation can escape along the
perpendicular outflow channels (e.g.,
\citealt{yorke2002,krumholz2005,kuiper2010,tanaka2016}).

In contrast to the indirect evidence, direct observational studies are
scarce, mainly because of two reasons: The clustered mode of high-mass
star formation and the typically large distances, making spatially
resolving such structures difficult.  Near-infrared interferometry has
imaged the warm inner dust disks around high-mass protostars
\citep{kraus2010,boley2013,boley2016}, showing that these are very
small ($<100$\,AU) and asymmetric structures.  However, this way one
can only study the innermost regions ($\leq 30$\,AU) whereas the total
gas and dust disk sizes are expected on scales of several 100 to
1000\,AU (e.g., \citealt{krumholz2007a,kuiper2011}). Additional
near-infrared evidence for disks around high-mass young stellar
objects stems from CO band-head emission (e.g.,
\citealt{bik2006,ilee2013}). So far, interferometric cm to submm
wavelength studies of high-mass disk candidates resulted in the
finding of larger-scale toroid structures of several 1000 to
$10^4$\,AU (e.g.,
\citealt{beuther2009c,beuther2013b,sandell2010,beltran2011,beltran2016}).
Almost all these objects are comparably high-mass and non-Keplerian
pseudo-disks, whereas Keplerian accretion-disk-like structures have so
far remained almost elusive to observations (well-known exceptions are
the intermediate-mass objects IRAS\,20126+4104 and AFGL490, e.g.,
\citealt{cesaroni2005,keto2010,schreyer2002}, or the famous source I
in Orion-KL, \citealt{greenhill2003,matthews2010}).  Just recently,
\citet{johnston2015} identified for the first time robust evidence of
a Keplerian disk around a high-mass protostar where in addition to the
central protostar the inner disk mass is also included to explain the
Keplerian motions.

The new Atacama Large Millimeter Array (ALMA) capabilities now allow
us to address many exciting issues. For example, simulations suggest
that the high-mass disks fragment at scales of about 300-500\,AU
(e.g., \citealt{krumholz2007a}). Submm interferometric observations at
690\,GHz (ALMA band 9) are inherently technically challenging because
of several reasons: Low precipitable water vapor (pwv) and a stable
atmosphere are needed to penetrate the atmosphere
properly. Furthermore, for phase and amplitude calibration, detections
of nearby quasars are needed, and the large collecting area of ALMA is
mandatory for that. In the past, only very few observations with the
Submillimeter Array (SMA) were conducted at these wavelengths (e.g.,
\citealt{beuther2006a,beuther2007c}), and ALMA is opening a new window
to related science. ALMA observations at 690\,GHz, with $>$1\,km
baselines, results in angular resolution elements $\leq 0.1''$,
corresponding to linear scales $\leq$200\,AU for our proposed
target. An additional unique feature of 690\,GHz observations is that
the dust continuum emission is very strong and can become optically
thick at the highest densities and smallest spatial scales, which
allows us to conduct spectral line absorption studies, ideally suited
to study the infall rates in the innermost regions around the
high-mass protostars (e.g., \citealt{wyrowski2012,beuther2013b}).\\

\paragraph{G351.77-0.54 (IRAS\,17233-3606):} This luminous
  high-mass star-forming region exhibits linear CH$_3$OH masers in
the northeast-southwest direction aligned with a CO outflow of similar
orientation \citep{walsh1998,leurini2009,leurini2013}. The first
  estimated luminosity of $\sim 10^5$\,L$_{\odot}$ is based on a
kinematic distance of 2.2\,kpc determined by \citet{norris1993} based
on a CH$_3$OH maser velocity of 1.2\,km\,s$^{-1}$. However, the
thermal line emission shows slightly lower values, usually below
$-$3\,km\,s$^{-1}$ (depending on the tracer, e.g.,
\citealt{leurini2011}). In the following, we use as a reference
$v_{\rm{lsr}}$ a value of $-3.63$\,km\,s$^{-1}$ measured for
CH$_3$CN$(12_2-11_2)v_8=1$ by \citet{leurini2011}. Since kinematic
distances around $v_{\rm{lsr}}$ velocities of 0 can only give rough
limits, smaller kinematic distances are derived, and
\citet{leurini2011} estimate an approximate distance of 1\,kpc. That
would then reduce the luminosity of the region to $\sim 1.7\times
10^4$\,L$_{\odot}$ \citep{leurini2011,leurini2013}. From large-scale
870\,$\mu$m continuum emission, \citet{leurini2011b} estimate a gas
mass reservoir of $\sim$664\,M$_{\odot}$ for a distance of 1\,kpc and
a temperature of 25\,K. For the proposed larger distance of 2.2\,kpc,
that would correspond to a 4.8 times larger mass. Both estimates
clearly show the large mass reservoir available for star formation in
this region. Using the Submillimeter Array (SMA) at 1.3\,mm with a
spatial resolution of $4.9''\times 1.8''$, \citet{leurini2011}
resolved a compact core with an estimated mass of 12\,M$_{\odot}$ at
the proposed 1\,kpc distance (again up to a factor 4.8 larger for
greater distance). \citet{beuther2009c} detected strong NH$_3$(4,4)
and (5,5) emission (excitation temperatures $\geq$200\,K) with a clear
velocity gradient approximately perpendicular to the (main)
outflow orientation, interpreted as evidence for rotational motion of
the core.  The rotating envelope has a projected diameter of $\sim
5''$ corresponding at the distance of $\sim$2.2\,kpc to an extent of
$\sim$11000\,AU, or at the closer distance of 1\,kpc to
$\sim$5000\,AU.  Since this structure encompasses six cm continuum
sources signposting fragmentation (\citealt{zapata2008}), several
smaller individual accretion disks and/or sub-sources may exist (see
clumpy sub-structure in Fig.~23 of \citealt{beuther2009c}).
Nevertheless, since the (multiple) outflows and rotation are not
strongly disturbed by the multiplicity, the region appears to be
dominated by one object, likely a real O-star progenitor.  After Orion
and SgrB2, it is one of the brightest line emission sources in the sky
and therefore a perfect ALMA high-frequency candidate.
%So far, no complementary data of the dust
%continuum emission, other spectral lines or the jet at high spatial
%resolution exist.

\begin{figure}[htb]
  \includegraphics[width=0.49\textwidth]{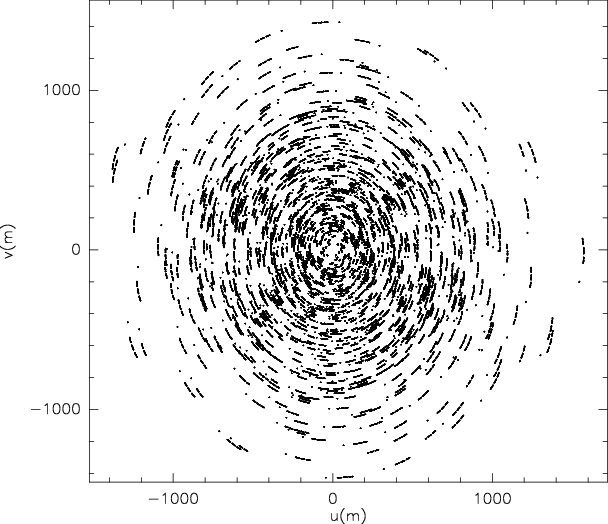}
  \caption{uv-coverage of these observations toward our target source.}
\label{uv-coverage} 
\end{figure} 

\begin{figure*}[htb]
 \includegraphics[width=0.49\textwidth]{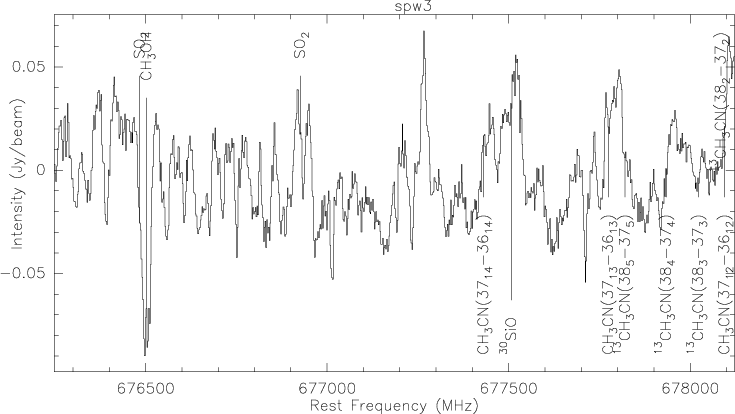}
 \includegraphics[width=0.49\textwidth]{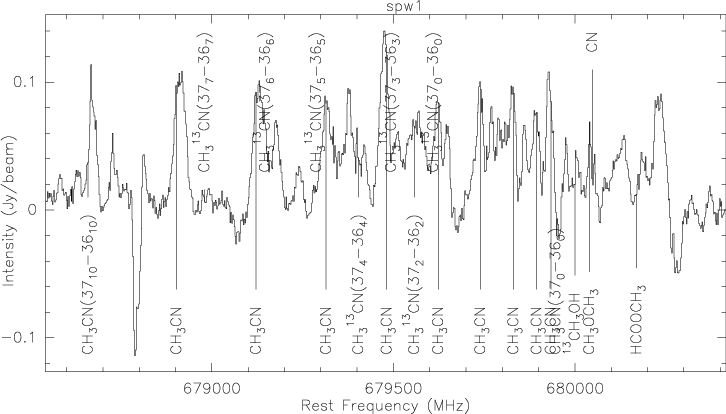}\\
 \includegraphics[width=0.49\textwidth]{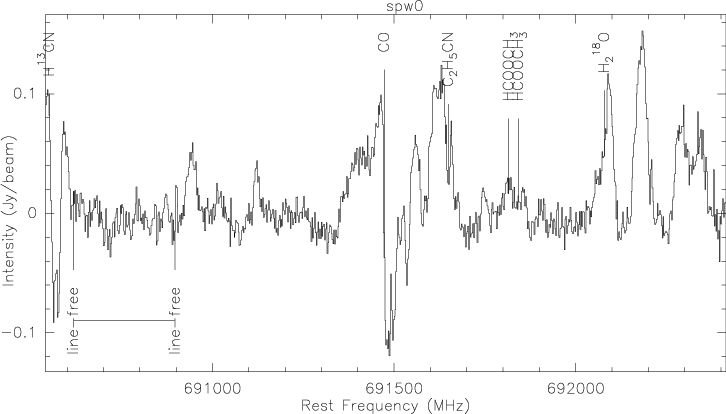}
 \includegraphics[width=0.49\textwidth]{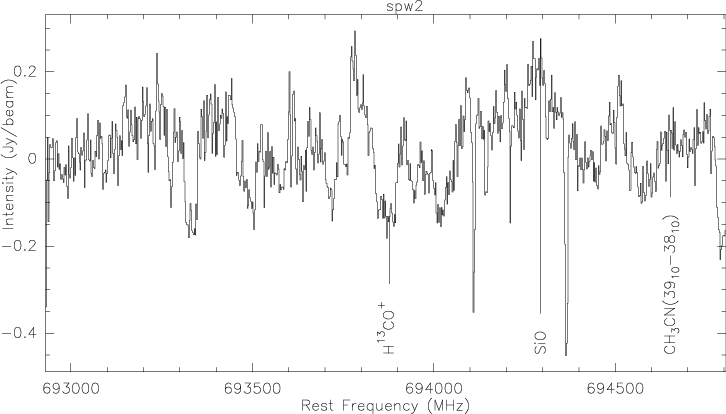}
 \caption{Spectra from the four spectral units (spw0 to spw3)
   extracted toward the main submm peak source \#1 (Table
   \ref{cont_table}) and averaging over $0.3''$ diameter. The
     spectra are ordered in frequency, not in spectral window units.
   The top two panels are the lower sideband and the bottom two panels
   the upper sideband. A few interesting lines as well as an almost
   line-free part in spw0 are marked.}
\label{spw} 
\end{figure*} 
                                                
\section{Observations} 
\label{obs}

The target region G351.77-0.54 was observed with ALMA in cycle 2 in
the 690\,GHz band with 42 antennas in the array. One noisy antenna and
very short baselines have been flagged because of outlying
amplitudes. This results in a baseline length coverage from 40\,m up
to almost $\sim$1.6\,km. The total duration of that one execution of
one scheduling block was 1\,h 40\,min with an on-target time of
35\,min. Because of the large number of antennas, even in such
comparably short time, ALMA achieves an excellent uv-coverage
(Fig.~\ref{uv-coverage}). This results in an almost circular beam (see
below) and allows us to recover spatial scales between approximately
$0.06''$ and $2.7''$. The phase center of the observations was
R.A.~(J2000.0) $17^{\rm{h}}26^{\rm{m}}42.568^{\rm{s}}$ and Dec.~
(J2000.0) $-36^{\circ}09'17.6''$.
% where the $v_{\rm{lsr}}$ was originally set to $+1.2$\,km\,s$^{-1}$. 
Bandpass calibration was conducted with observations of J1924-2914,
and the absolute flux was calibrated with observations of Pallas. For
the gain phase and amplitude calibration, the regularly interleaved
observations of J1717-3342 were used. The quasar was visited every
7\,min for 1\,min integration time, allowing to track the phase
variability and associated calibration well. To check the solutions in
this difficult to observed frequency range, another quasar J1745-2900
was observed as well. Imaging that quasar results in the expected very
good image of a point source at the phase center, confirming the good
calibration. The receivers were tuned to an LO frequency of
684.405\,GHz with four spectral windows distributed evenly in the
lower and upper sideband.The width and spectral resolution of each
spectral unit was 1.875\,GHz and 488\,kHz, respectively (corresponding
in velocity space to $\sim$815\,km\,s$^{-1}$ and
$\sim$0.21\,km\,s$^{-1}$). The central frequencies of the four windows
are 677.185, 679.479, 691.478, and 693.867\,GHz, respectively
(Fig.~\ref{spw}). The primary beam for ALMA at this wavelength is
$\sim 9.2''$. The calibration was conducted in CASA version 4.3.1 with
the ALMA-delivered reduction script. In a following step we tested
self-calibration (with solution intervals of 30\,sec and shorter), but
since that did not improve the data, we used the non-self-calibrated
data.

\begin{figure*}[htb]
  \includegraphics[width=0.99\textwidth]{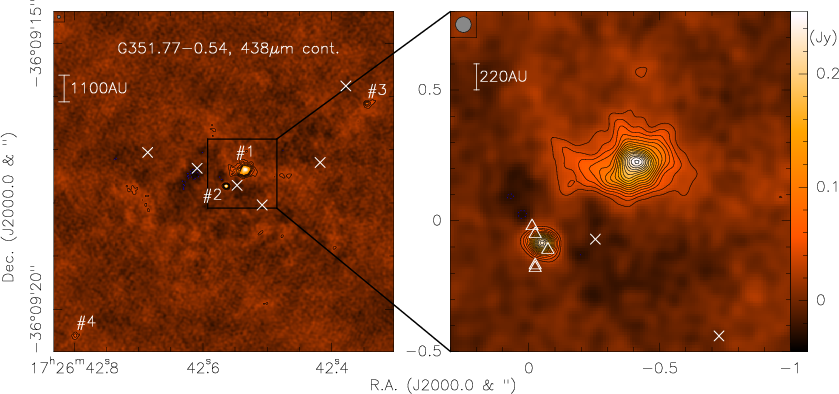}
  \caption{438\,$\mu$m continuum image of G351.77-0.54. The left panel
    shows a larger area encompassing the two central sources as well
    as two additional submm sources in the south-east and north-west
    (all marked). The white crosses mark the cm continuum sources by
    \citet{zapata2008}. The right panel then shows a zoom into the
    center with the triangles presenting additionally the class II
    CH$_3$OH masers from \citet{walsh1998}. The contour levels start
    at $4\sigma$ and continue in $2\sigma$ steps with the $1\sigma$
    level of 7\,mJy\,beam$^{-1}$. The synthesized beam of $0.06''$ and
    a scale-bar for an assumed distance of 2.2\,kpc are presented as
    well.}
\label{cont} 
\end{figure*} 

Imaging and further data processing was also conducted in CASA.  As
visible in Fig.~\ref{spw}, there are very few line-free parts of the
spectra, and creating a pure continuum image is difficult. Therefore,
for the continuum data, we avoided the strongest line emission
features (spw1 for the main CH$_3$CN emission, and the strongest
features in spw0 -- 690.537 to 690.634\,GHz, 691.357 to 691.870\,GHz
and 692.042 to 692.417\,GHz), and collapsed the remaining spectral
coverage into a continuum dataset. While these data provide a good
signal-to-noise ratio they are still line-contaminated. To test how
large that effect is, we created an additional continuum dataset with
a small spectral band in spw0 from 690617\,MHz to 690897\,MHz that
appears almost line-free (Fig.~\ref{spw}). Comparing the flux values
in the corresponding two images, we find that the continuum fluxes
largely agree within 10\% increasing at most to 20\%
difference. Hence, line contamination does not affect our continuum
data much. Applying a robust weighting of -2 for the continuum data
and a robust weighting of 0.5 to increase the signal-to-noise ratio
for the line data, we achieve spatial resolution elements of $0.06''$
and $0.07''$ for the continuum and line data, respectively. The
$1\sigma$ rms of the 438\,$\mu$m continuum and spectral line maps at
0.5\,km\,s$^{-1}$ resolution are 7\,mJy\,beam$^{-1}$ and
30\,mJy\,beam$^{-1}$ (corresponding to brightness temperature
sensitivities of $\sim$5.1\,K and $\sim$16.0\,K), respectively.

The study presented here focuses on the analysis of the submm
continuum as well as the spectral line emission from CH$_3$CN (as
dense gas tracer) and CO (as outflow tracer).

\section{Results}

\subsection{Submm continuum emission at 438\,$\mu$m}
\label{submm_cont}

Figure \ref{cont} presents the 438\,$\mu$m continuum image of the
G351.77-0.54 region. The first impression one gets from this image is
that on scales of several 1000\,AU we find only a small degree of
fragmentation. In the remainder of the paper, we will always state the
quantitative parameters for both distance estimates of 2.2 and
1.0\,kpc (see Introduction), where the 1.0\,kpc parameter will be
stated in brackets. While within approximately the inner 500(250)\,AU
the region fragments into two cores (\#1 and \#2), outside of that we
have several 1000\,AU scales without any further submm core, and then
we find the two additional cores \#3 and \#4. Assuming optically thin
dust emission at 200\,K (see more details below), our $3\sigma$
detection limit of 21\,mJy\,beam$^{-1}$ corresponds to mass and column
density sensitivities of $\sim$0.004($\sim$0.0009)\,M$_{\odot}$ and
$\sim 5\times 10^{23}$\,cm$^{-2}$, respectively. While we can
obviously miss some low-column density cores, assuming a distribution
of column densities it seems unlikely that we are missing a whole
population of lower-mass cores in this region. Figure \ref{cont} also
shows the five cm continuum sources reported by \citet{zapata2008}. We
also present the CH$_3$OH class II maser positions from
\citet{walsh1998} in Figure \ref{cont}, and they seem to be associated
with source \#2. However, the absolute maser positions of this dataset
are only certain within $1''$, hence, we cannot associate them
unambiguously with any of the source and hence refrain from any
further speculation about them. Since none of them has a submm
counterpart, the protostellar nature of these emission features is
unclear, and it may be that the cm emission stems from protostellar
jets in the region (see also section \ref{outflow}).

Zooming into the central two cores (Fig.~\ref{cont} right panel) we
clearly resolve the region at $0.06''$ resolution, corresponding to
linear resolution elements of 132(60)\,AU. While source \#2 is almost
unresolved by these observations, the main source \#1 is clearly
extended in the east-west direction, approximately perpendicular to
the main outflows reported in \citet{leurini2011}. As will be shown
below, \#1 is extremely strong in line emission whereas \#2 is
weaker. Without kinematic evidence we cannot provide firm proof of
this, but it seems likely that \#1 and \#2 could form a high-mass
binary system. It seems likely that the central part (essentially the
central beam of 132(60)\,AU diameter) is optically thick in the
438\,$\mu$m dust continuum emission. This is partially based on the
high Planck brightness temperature of the central beam towards source
\#1 of approximately 201\,K (see section \ref{temp_struc}). For a
protostar of luminosity of $\sim2-10\times 10^4$\,L$_{\odot}$ and
assuming a typical dust temperature distribution $T\propto r^{-0.4}$,
temperatures on the order of 200\,K at 100\,AU distance are expected,
agreeing reasonably well with the high brightness temperature and
moderate optical depth. At such comparably high optical depth of the
submm continuum emission, we cannot unambiguously distinguish whether
source \#1 is a single source or whether multiplicity is hidden by
optical depth. As we discuss in section \ref{temp_struc}, the optical
depth should decrease rapidly with radius. It is worth noting also
that the CH$_3$CN$(37_k-36_k)$ emission (see section \ref{dense})
appears to follow an arc or ring centered on the continuum maximum
which also suggests dust obscuration towards the peak of source
\#1. Our mass estimates given below based on the optically thin
assumption are thus underestimates though likely not by a large
factor.

\begin{figure}[htb]
  \includegraphics[width=0.49\textwidth]{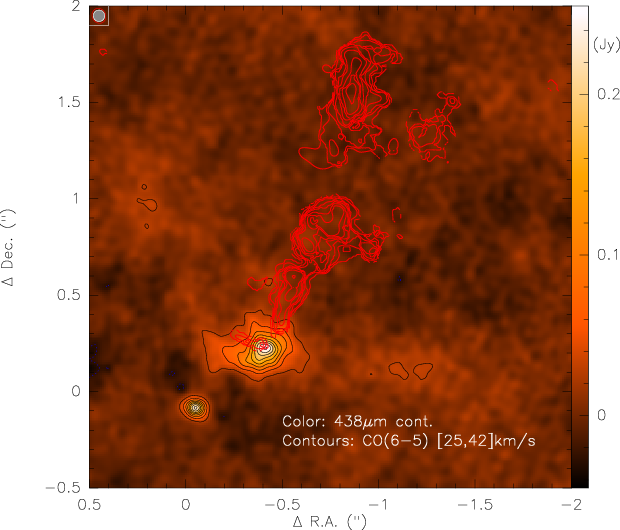}
  \caption{438\,$\mu$m continuum image with CO(6--5) red-shifted
    emission in red contours. The continuum contours are in $4\sigma$
    steps ($1\sigma=7$\,mJy\,beam$^{-1}$. The blue-dotted contours
    show the corresponding negative features. The CO emission was
    integrated from 25 to 42\,km\,s$^{-1}$ and is shown in contours
    from 15 to 95\% (step 10\%) of the peak emission.  The center
    coordinates are those of the phase center presented in section
    \ref{obs}.}
\label{co} 
\end{figure} 

\begin{figure}[htb]
  \includegraphics[width=0.49\textwidth]{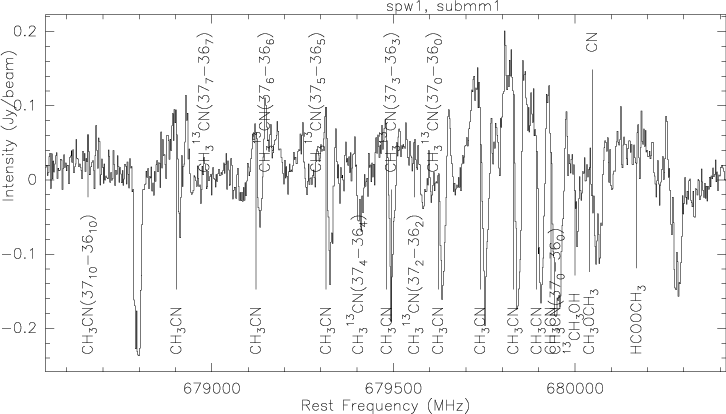}
  \includegraphics[width=0.49\textwidth]{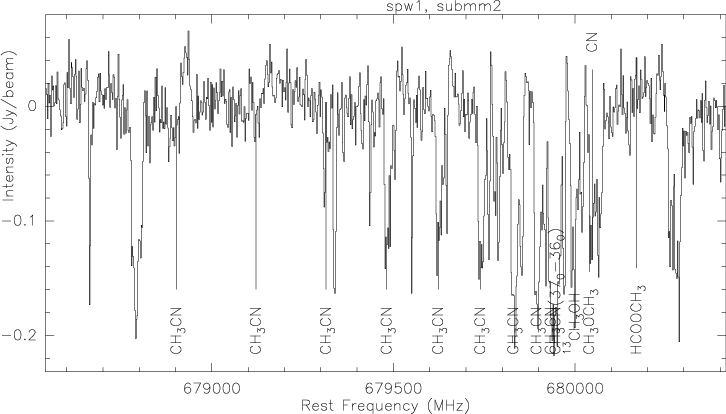}
  \includegraphics[width=0.49\textwidth]{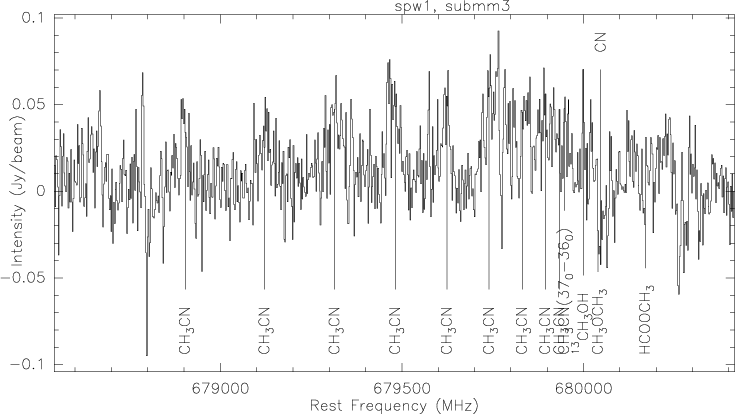}
  \includegraphics[width=0.49\textwidth]{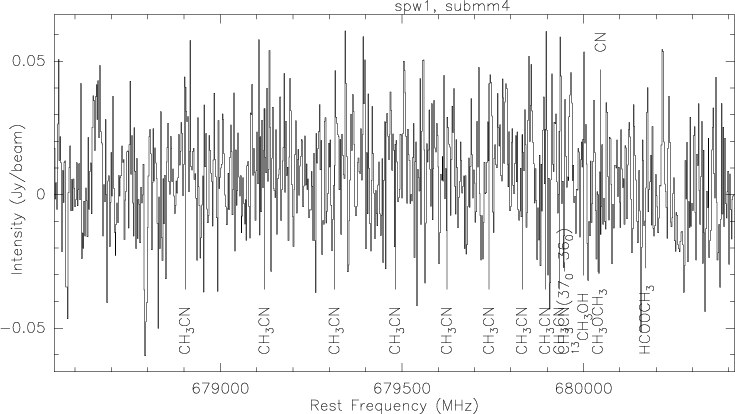}
  \caption{Spectra extracted toward all four submm peaks in spectral
    win spw1 averaged over $0.06''$ diameter (units Jy\,beam$^{-1}$).}
\label{submm2-4} 
\end{figure} 

The submm continuum emission can also be employed to estimate column
densities and masses of the four submm sources. Table \ref{cont_table}
presents the peak positions, the corresponding peak flux densities and
Planck brightness temperatures as well as the total fluxes integrated
within the $4\sigma$ contour levels. Although the highest column
density peak positions have increased optical depth, offset from the
peak position most of the dust continuum emission should still be
optically thin. Therefore, assuming optically thin emission from
thermally emitting dust grains we can estimate lower limits of the
column densities and masses following the outline of
\citet{hildebrand1983} or \citet{schuller2009}. For these estimates we
use a dust opacity $\kappa_{438\mu\rm{m}}=6.3$\,cm$^2$\,g$^{-1}$,
extrapolated for densities of $10^6$\,cm$^{-3}$ from
\citet{ossenkopf1994}, and a conservative hot core temperature of
200\,K is used for all calculations here (measured continuum peak
brightness temperatures are $\sim$201\,K, see also temperature
discussion below, section \ref{temp_struc}).  Furthermore, we use a
gas-to-dust mass ratio of 150 \citep{draine2011}. The derived column
densities are all above $10^{24}$\,cm$^{-2}$ showing that we are
dealing with extremely high extinction regions in excess of
1000\,magnitudes of visual extinction. In comparison, the mass
estimates are rather low around 0.49(0.1)\,M$_{\odot}$ assuming a
distance of 2.2(1.0)\,kpc. The main reason for that is the large
amount of spatial filtering because of the missing baselines below
40\,m (section \ref{obs}). Although no single-dish 438\,$\mu$m
measurement exists to exactly estimate the amount of missing flux, a
comparison of the single-dish mass reservoir estimated by
\citet{leurini2011b} of 664\,M$_{\odot}$ (@1\,kpc) indicates that more
than 99\% of the extended flux is filtered out by our extended
baseline ALMA data. This is expected since the single-dish data trace
the larger-scale less dense envelope structure. Even on smaller scales
($4.9''\times 1.8''$), \citet{leurini2011} measure a peak flux of
2.09\,Jy\,beam$^{-1}$ at 1.3\,mm wavelength. Assuming a spectrum
$\propto \nu^{3.5}$ in the Rayleigh-Jeans limit (with
$3.5=2+\beta$ and the dust opacity index $\beta = 1.5$) this would
correspond to $\sim 98$\,Jy\,beam at our wavelength of
438\,$\mu$m. Even in the optically thick case with $beta =0$ a flux
density of 18.9\,Jy\,beam$^{-1}$ would be expected at 438\,$\mu$m. So,
even on these smaller scales, our ALMA data recover only a small
fraction of the core flux. Additionally, at 438\,$\mu$m the assumption
of optically thin emission breaks down (also visible in the absorption
spectra presented in the section \ref{dense}), and one underestimates
the masses and column densities as well. The 0.49\,M$_{\odot}$
measured toward source \#1 may be considered as a lower limit to the
mass associated with that source (see section
\ref{dense}). Nevertheless, as outlined in the introduction, the
region has a very large mass reservoir on the order of
1000\,M$_{\odot}$ capable of forming high-mass stars.

\begin{table*}[htb]
\caption{Continuum 438\,$\mu$m parameters}
\begin{tabular}{lrrrrrrr}
\hline
\hline
Name & R.A.~(J2000.0) & Dec.~(J2000.0)& $S_{\rm{peak}}$ & $T_{\rm{bright}}$ & $S_{\rm{int}}^a$ & $N_{\rm{H}_2}^b$ & $M^b$ \\
     & (h:m:sec) & ($^{\circ}$:$'$:$''$) & (Jy/beam) & (K) & (Jy) & ($\times 10^{24}$cm$^{-2}$) & $M_{\odot}$ \\
\hline
\#1 & 17:26:42.534 & -36:09:17.38 & 0.255 & 201 & 2.33 & 6.2 & 0.49 \\
\#2 & 17:26:42.564 & -36:09:17.69 & 0.231 & 183 & 0.364 & 5.6 & 0.08 \\
\#3 & 17:26:42.345 & -36:09:16.13 & 0.082 & 75  & 0.256 & 2.0 & 0.05 \\
\#4 & 17:26:42.797 & -36:09:20.50 & 0.048 & 49  & 0.125 & 1.2 & 0.03 \\
\hline                                          
\hline                                          
\end{tabular}                                  
{\footnotesize
  ~\\
  $^a$ The integrated flux densities are measured within the $4\sigma$ contours.\\
  $^b$ Dust temperatures are assumed to 200\,K (see main text). The masses are calculated for an assumed distance of 2.2\,kpc and are lower limits because of optical depth effects.
}
\label{cont_table}                                   
\end{table*}                                     

\subsection{Outflow emission}
\label{outflow}

With the shortest baselines around 40\,m (corresponding to spatial
scales of $\sim$2.7$''$, see section \ref{obs}), it is extremely
difficult to identity any coherent large-scale structure in this
dataset. Nevertheless, we investigated the CO(6--5) observations
channel by channel to search for potential high-velocity gas from a
molecular outflow. While we do not find any such signature in the
blue-shifted gas, we do clearly identify a red-shifted outflow lobe in
the velocity regime from 25 to 42\,km\,s$^{-1}$. Figure \ref{co}
presents the integrated CO(6--5) emission in that velocity regime
overlaid on the dust continuum emission. Although the blue-shifted
counterpart is missing - most likely because the blue-shifted gas may
be more diffusely distributed and thus filtered out - the almost
cone-like red-shifted emission emanating from the main submm peak \#1
in north-northwestern direction exhibits the clear structural shape of
a molecular outflow. Although \citet{leurini2011} identify multiple
outflows in the region, one of them (OF1 in their nomenclature)
appears consistent with the main outflow we find here. The
larger-scale northeast-southwest outflow in \citet{leurini2011} is not
recovered in our new high-resolution data here.

Additionally, almost perpendicular to the main outflow structure, we
find another much smaller red-shifted emission structure emanating
from the main submm peak \#1 in approximately east-northeastern
direction. This feature may stem from a second outflow driven from
within the main source \#1. As outlined in section \ref{submm_cont},
source \#1 can easily host a secondary source, either at unresolved
spatial scales or even hidden by the high optical depth of the
continuum emission.

\begin{figure}[htb]
  \includegraphics[width=0.49\textwidth]{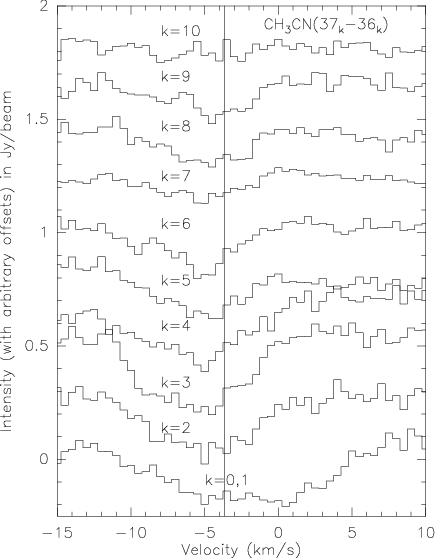}
  \caption{CH$_3$CN$(37_k-36_k)$ spectra extracted toward the submm
    continuum peak position \#1 from all $k$-components. The approximate
    $v_{\rm{lsr}}$ is marked by a vertical line, and the spectra are
    shifted in the y-axis for clarity.}
\label{peak_spectra} 
\end{figure} 

\begin{figure}[htb]
  \includegraphics[width=0.49\textwidth]{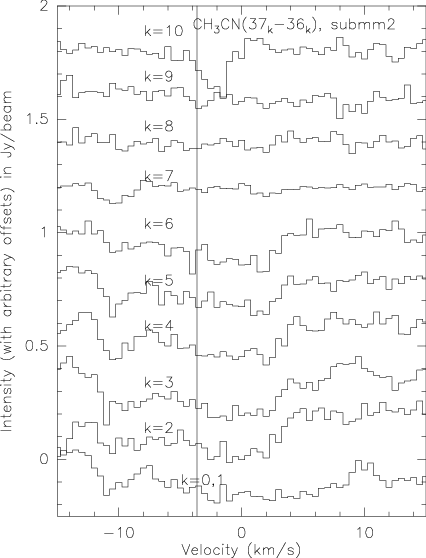}
  \caption{CH$_3$CN$(37_k-36_k)$ spectra extracted toward the submm
    continuum position \#2 from all $k$-components. The approximate
    $v_{\rm{lsr}}$ is marked by a vertical line, and the spectra are
    shifted in the y-axis for clarity.}
\label{peak_submm2} 
\end{figure} 

\subsection{Dense gas spectral line emission}
\label{dense}

As shown in Fig.~\ref{spw}, the region is a typical hot molecular core
with extremely rich spectral line emission. In this paper, we focus on
the dense gas and dust emission. Therefore, in the following, we
concentrate on the hot core and dense molecular gas tracer methyl
cyanide which is observed here in its main isotopologue CH$_3$CN and
its rarer isotopologue CH$_3^{13}$CN in the transitions $(37_k-36_k)$
with $k$-levels from 0 to 10. CH$_3$CN traces the dense gas and is a
well-known tool to study the dynamics at the center of high-mass
star-forming regions (e.g., \citealt{cesaroni2007}). Furthermore, the
lines of the rotational $k-$ladder can be used as a thermometer (e.g.,
\citealt{loren1984}). Table \ref{lines} presents the frequencies and
lower level energies $E_l/k$ of the respective transitions.

\begin{table}[htb]
\caption{Parameters of main CH$_3$CN$(37_k-36_k)$ lines in spectral
  window 1 (spw1)}
\begin{tabular}{lrr}
\hline
\hline
Line & Freq. & $E_l/k$ \\
     & (GHz) & (K) \\
\hline
CH$_3$CN$(37_0-36_0)$ & 679.947 & 588 \\
CH$_3$CN$(37_1-36_1)$ & 679.934 & 595 \\
CH$_3$CN$(37_2-36_2)$ & 679.895 & 616 \\
CH$_3$CN$(37_3-36_3)$ & 679.831 & 652 \\
CH$_3$CN$(37_4-36_4)$ & 679.740 & 701 \\
CH$_3$CN$(37_5-36_5)$ & 679.624 & 766 \\
CH$_3$CN$(37_6-36_6)$ & 679.482 & 844 \\
CH$_3$CN$(37_7-36_7)$ & 679.315 & 937 \\
CH$_3$CN$(37_8-36_8)$ & 679.122 & 1043 \\
CH$_3$CN$(37_9-36_9)$ & 678.903 & 1165 \\
CH$_3$CN$(37_{10}-36_{10})$ & 678.659 & 1300 \\
CH$_3^{13}$CN$(37_0-36_0)$ & 679.610 & 587 \\
CH$_3^{13}$CN$(37_1-36_1)$ & 679.597 & 595 \\
CH$_3^{13}$CN$(37_2-36_2)$ & 679.558 & 616 \\
CH$_3^{13}$CN$(37_3-36_3)$ & 679.494 & 652 \\
CH$_3^{13}$CN$(37_4-36_4)$ & 679.404 & 702 \\
CH$_3^{13}$CN$(37_5-36_5)$ & 679.288 & 766 \\
CH$_3^{13}$CN$(37_6-36_6)$ & 679.147 & 844 \\
CH$_3^{13}$CN$(37_7-36_7)$ & 678.980 & 937 \\
\hline                                          
\hline                                          
\end{tabular}                                  
\label{lines}                                   
\end{table}                                     

The $E_l/k$ values between 588 and 1300\,K show that we are tracing a
very warm gas component. Furthermore, it is interesting to note that
we do see pure emission only toward the extended parts of submm source
\#1 whereas toward the regions of peak emission of the central two
submm peaks \#1 and \#2 the spectra are dominated by absorption
(Figs.~\ref{submm2-4}, \ref{peak_spectra} and \ref{peak_submm2}). In
comparison to Fig.~\ref{lines} where the spectra toward the region of
the main peak \#1 are shown after averaging over $0.3''$ diameter to
show the general absorption and emission features, Fig.~\ref{submm2-4}
presents the corresponding spectra in spw1 toward the regions of all
four continuum peaks with a smaller averaging diameter of $0.06''$
(units Jy\,beam$^{-1}$). The secondary source \#2 is clearly an
absorbing source, while source \#3 shows weak emission and source \#4
exhibits mainly a noise spectrum.

\subsubsection{Kinematics of source \#1}

\begin{figure*}[htb]
  \includegraphics[width=0.99\textwidth]{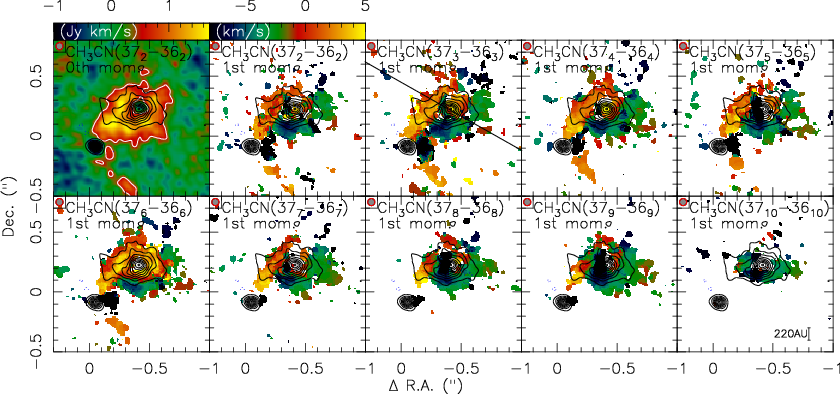}
  \caption{CH$_3$CN($37_k-36_k$) 1st moment maps of the the $k$-levels
    2 to 10 toward the central sources \#1 and \#2. The top-left panel
    shows the corresponding integrated 0th moment map of the
    CH$_3$CN($37_2-36_2$) transition (integrated from $-13$ to
    +10\,km\,s$^{-1}$). The white contour outlines the 3$\sigma$
      level of 0.51\,Jy\,km\,s$^{-1}$. The other panels all show
    intensity-weighted velocity maps, and wedges are shown above the
    $k=2$ panels. While the $k=0$ map is done without any clipping,
    the first moment maps are clipped at a 4$\sigma$ level. The black
    contours present the 438\,$\mu$m continuum emission in $4\sigma$
    steps ($1\sigma=7$\,mJy\,beam$^{-1}$).  The synthesized beams are
    shown for all panels, and a scale-bar (@2.2\,kpc) is presented in
    the bottom-right panel. The diagonal line in the top-middle panel
    of CH$_3$CN$(37_3-36_3)$ marks the axis for the position-velocity
    diagrams shown in Fig.~\ref{pv}.}
\label{mom1} 
\end{figure*}

The absorption spectra toward the main source \#1 in
Fig.~\ref{peak_spectra} are dominated by blue-shifted absorption (the
blended $k=0,1$ components are too broad to isolate blue- and
red-shifted absorption properly). Hence, the dense gas along the line
of sight of that main peak appears to be dominated by outflowing gas
kinematics. Interestingly, the CH$_3$CN$(37_{10}-36_{10}$) component
toward the main submm peak \#1 exhibits neither absorption nor
emission but only a flat noise spectrum. Since this $k=10$ component
shows clear emission in the immediate environment (see below and
Fig.~\ref{mom1}), sensitivity seems unlikely to be the reason for that
non-detection. The more reasonable explanation is high optical depth
of the continuum. With an excitation temperature $E_l/k=1300$\,K, the
main excitation region of the $k=10$ component may stem from inside
the $\tau =1$ surface of the 438\,$\mu$m continuum emission and can
thus be veiled by the continuum.

A different way to analyze the CH$_3$CN data is via spatially imaging
the data and discussing the kinematics via moment and
position-velocity analysis. Fig.~\ref{mom1} presents the 1st moment
maps (intensity-weighted peak velocities) of the CH$_3$CN$(37_k-36_k)$
data for the $k$-components $2-10$. ($k=0,1$ are blended). The Figure
also shows for comparison the integrated emission of the $k=2$
component. This integrated intensity map exhibits a ring-like
structure around the main peak with an approximate diameter of $\sim
0.28''$, however, this does not imply a ring-like CH$_3$CN
distribution but is certainly caused by absorption toward the
center. Although this ring-like structure is not entirely smooth,
intensity variations within that structure usually do not exceed
$3\sigma$. Hence, we refrain from further interpretation of such
CH$_3$CN fluctuations.

\begin{figure*}[htb]
  \includegraphics[width=0.99\textwidth]{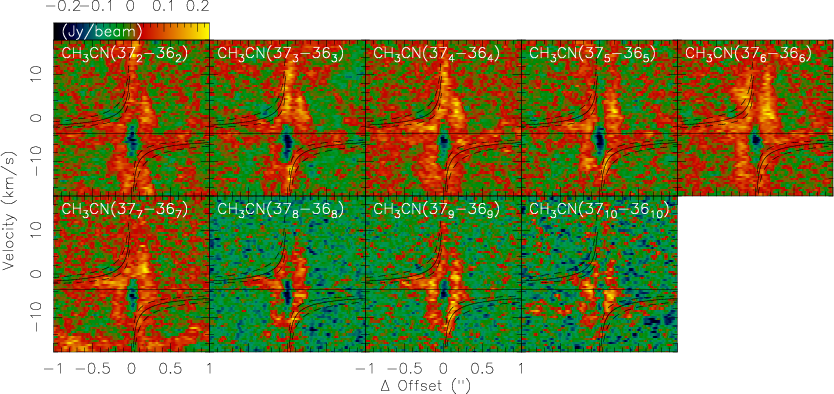}
  \caption{Position-velocity diagrams for the nine
    CH$_3$CN($37_k-36_k$) transitions toward source \#1 along the axis
    marked in the top-middle panel of Fig.~\ref{mom1}. The vertical
    line shows the approximate velocity of rest and the two full-line
    curves present a Keplerian rotation curve for a
    10(4.5)\,M$_{\odot}$ star at distances of 2.2(1)\,kpc,
    respectively. The dashed curves correspond to masses of 5(2.3) and
    20(9)\,M$_{\odot}$ (inner and outer curves, respectively) at
    2.2(1)\,kpc distance.}
\label{pv} 
\end{figure*} 

As shown in section \ref{outflow}, the CO(6--5) data reveal a clear
outflow in the north-northwest direction and additional high-velocity
gas almost perpendicular to this outflow. Based on the main outflow
structure and also indicated by the 1st moment maps, we identify the
axis for the rotational structure perpendicular to the main outflow
and mark that axis in the CH$_3$CN$(37_3-36_3)$ panel in Figure
\ref{mom1}. While this axis is clearly perpendicular to the outflow
presented in section \ref{outflow} and figure \ref{co},
\citet{leurini2011} showed multiple outflows emanating from this
region. Hence, the CH$_3$CN velocity structure is likely also
influenced by additional outflow kinematics. Along this axis marked in
Fig.~\ref{mom1} we present position-velocity (pv) cuts for the
$k=2...10$ components of CH$_3$CN$(37_k-36_k)$ in Fig.~\ref{pv}. These
pv-cuts exhibit several interesting features. First, one can clearly
identifies the absorption structure toward the continuum peak
dominated by blue-shifted absorption, again reinforcing that we do not
just see the dense rotating structure in the CH$_3$CN gas but also
contributions from the outflow. Furthermore, the pv diagrams partly
show structures reminiscent of Keplerian rotation with the highest
velocities close to the center and lower velocities further away from
the center. To illustrate this, the pv-diagram also shows the expected
signature for a Keplerian disk around a 10(4.5)\,M$_{\odot}$ star at
2.2(1)\,kpc, respectively. In particular for the $k=2...6$ components,
the pv diagram reflects this structure. However, there is a caveat to
this interpretation: A real Keplerian pv diagram would only show the
Keplerian structures as indicated by the lines in Fig.~\ref{pv}, but
we see similar structures in the two opposing quadrants of these
pv-diagrams which are untypical for Keplerian disks. However,
\citet{seifried2016} have recently shown that, depending on the
inclination, emission from disks can also appear in these other
quadrants of the pv-diagrams. Nevertheless, it appears likely that the
CH$_3$CN emission includes additional gas components that are not
represented by a pure Keplerian rotation model, e.g., emission from
the infalling envelope and the outflows (see section
\ref{discussion_disk} for further discussion).

\subsubsection{Infall motions toward source \#2}

In contrast to the main peak, the absorption spectra for the secondary
peak \#2 (see Fig.~\ref{peak_submm2}) exhibit absorption on the blue-
and red-shifted side of the spectrum. This can be seen in particular
for the $k-$components 2 to 6 (again the 0,1 components are blended),
while the K-components 7 to 9 are showing mainly noise. Interestingly,
the $k=10$ component of CH$_3$CN in Fig.~\ref{peak_submm2} again shows
a clear and pronounced red-shifted absorption feature. However, since
the $k$- 7 to 9 components do not show this behavior, we caution that
this absorption feature near the $k=10$ component may be a line-blend
from an unidentified species. Nevertheless, based on the $k=2-6$
CH$_3$CN$(37_k-36_k)$ lines, infall and outflow motion toward/from
submm source \#2 appears the most likely kinematic interpretation of
the data. 

We cannot spatially resolve any kinematics but just analyze the
kinematics along the line of sight. In addition to the blue-shifted
absorption likely stemming from outflowing gas, the red-shifted
absorption spectra in the $k=2...6$ components are indicative of
infall motions toward that core. If we assume spherical symmetry, we
can estimate the mass infall rate with $\dot{M}_{\rm{in}} = 4\pi r^2
\rho v_{\rm{in}}$ where $\dot{M}_{\rm{in}}$ and $v_{\rm{in}}$ are the
infall rate and infall velocity, and $r$ and $\rho$ the core radius
and density (see also \citealt{qiu2011,beuther2013b}). Since the
infall rate scales linearly with the distance, the following estimate
is only conducted for the distance of 2.2\,kpc and can be scaled to
1.0\,kpc. We estimate the radius to $r\sim 80$\,AU which is half our
spatial resolution limit in the line emission of $0.07''$. For the
density estimate, we $\rho\sim 2.3\times 10^9$\,cm$^{-3}$ assuming
that the peak column density (Table \ref{cont_table}) is smoothly
distributed along the line-of-sight at our spatial-resolution limit
(i.e., 160\,AU at 2.2\,kpc). The infall velocity can be estimated to
$\sim$7.1\,km\,s$^{-1}$, corresponding to the difference between the
most red-shifted absorption at $\sim +3.5$\,km\,s$^{-1}$, and the
$v_{\rm{lsr}}\sim -3.6$\,km\,s$^{-1}$. Using these physical
parameters, we can estimate approximate mass infall rates toward submm
source \#2 of $\dot{M}_{\rm{in}} \sim 1.6\times
10^{-3}$\,M$_{\odot}$\,yr$^{-1}$. At a distance of 1\,kpc, that infall
rate is a factor 2.2 lower.
%GREG> say '4*pi*(80*1.495985e13)**2*2.3e9*7.1e5*(2*1.6725e-27)/1.989e30*(60*60*24*365)' 
%1.5588288918642E-03 

Since collapse motions in star formation are typically associated with
outflow and rotational dynamics, we hypothesize a 2D disk-like
geometry to refine the infall rate estimate following
\citet{beuther2013b}: If the accretion follows along a flattened
disk-like structure with a solid angle $\Omega$ (instead of the $4\pi$
in the spherical case), the corresponding disk infall rate
$\dot{M}_{\rm{disk,in}}$ scales approximately like
$\dot{M}_{\rm{disk,in}}=\frac{\Omega}{4\pi}\times
\dot{M}_{\rm{in}}$. Following
\citet{kuiper2012,kuiper2015,kuiper2016}, an outflow covers
approximately an opening angle of $120^{\circ}$ while the disk covers
$\sim$60$^{\circ}$ (both to be doubled to take into account the
north-south symmetry). Because opening angles do not scale linearly
with the surface elements, full integration results in approximately
50\% or $\sim 2\pi$ being covered by the disk-like structure. Taking
these arguments into account, we estimate approximate disk infall
rates around $\dot{M}_{\rm{disk,in}}\sim 0.8\times 10^{-3}$
\,M$_{\odot}$\,yr$^{-1}$ at 2.2\,kpc distance, or again a factor 2.2
lower at 1.0\,kpc. Although strictly speaking these are disk infall
rates and not accretion rates, one can use them as upper limit for the
mean actual accretion rates (due to variable accretion from disk to
star, the actual accretion rate can be much higher than the large
scale infall rate on short timescales), and they are fully consistent
with expected accretion rates for high-mass star formation (e.g.,
\citealt{krumholz2007b, beuther2006b, zinnecker2007, mckee2007,
  kuiper2010,tan2014}).

\subsection{Temperature structure}
\label{temp_struc}

\begin{figure}[htb]
\begin{center}
  \includegraphics[width=0.49\textwidth]{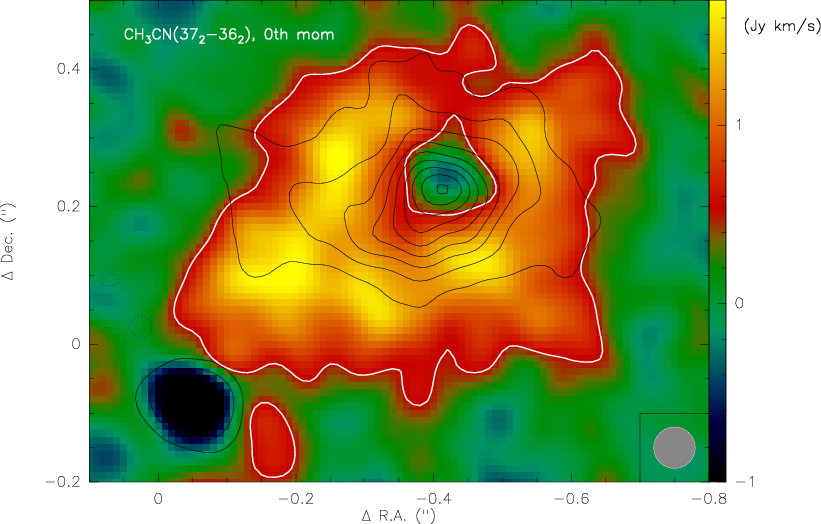}
  \includegraphics[width=0.43\textwidth]{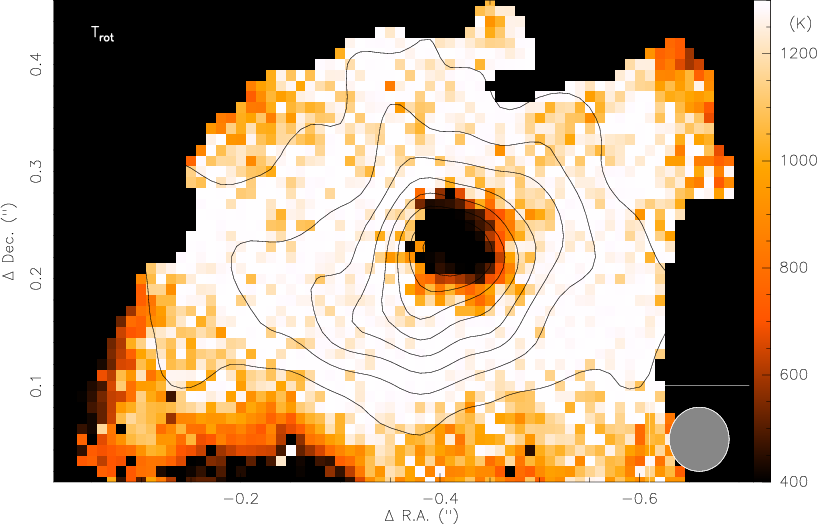}
  \includegraphics[width=0.42\textwidth]{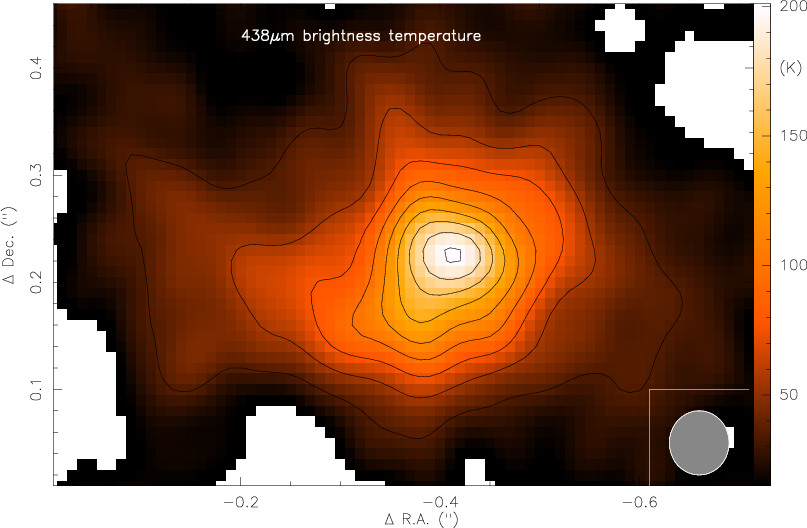}
\end{center}
\caption{The top panel shows for source \#1 the moment 0 map for
  CH$_3$CN$(37_2-36_2)$ integrated from -13 to +10\,km\,s$^{-1}$. The
  white contour outlines the 3$\sigma$ level of
  0.51\,Jy\,km\,s$^{-1}$. The middle panel presents the rotational
  temperature map derived with XCLASS from the full
  CH$_3$CN$(37_k-36_k)$ ladder ($k=0...10$). Temperatures outside the
  CH$_3$CN$(37_2-36_2)$ 3$\sigma$ level are blanked. For comparison,
  the bottom panel presents the 438\,$\mu$m continuum emission
  converted to Planck brightness temperatures. The black contours show
  in all panels the 438\,$\mu$m continuum emission in $4\sigma$ steps
  of 28\,mJy\,beam$^{-1}$. The beam is shown at the bottom-right of
  each panel.}
\label{temp} 
\end{figure}

The CH$_3$CN$(37_k-36_k)$ $k-$ladder allows a detailed derivation of
the gas temperature of the central rotating core/disk structure 
  associated with source \#1. With energy levels between 588 and
1300\,K (Table \ref{lines}) we are tracing the hot component of the
gas. Since the same spectral setup also covers the
CH$_3^{13}$CN$(37_k-36_k)$ isotopologue (Fig.~\ref{spw} and Table
\ref{lines}), high optical depth can also be taken into account. In
addition to this, at the given very high spatial resolution, we do not
get just an average temperature, but we can derive the temperature
structure pixel by pixel over the full extent of the CH$_3$CN
emission. Figure \ref{temp} shows in the top panel the integrated 0th
moment map of the CH$_3$CN$(37_2-36_2)$ to outline the extent of the
CH$_3$CN emission in comparison to the submm continuum.

For the pixel-by-pixel fitting of the CH$_3$CN$(37_k-36_k)$ data we
used the XCLASS (eXtended CASA Line Analysis Software Suite) tool
implemented in CASA \citep{moeller2015}. This tool models the data by
solving the radiative transfer equation for an isothermal homogeneous
object in local thermodynamic equilibrium (LTE) in one dimension,
relying on the molecular databases VAMDC and CDMS
(http://www.vamdc.org and \citealt{mueller2001}). It furthermore uses
the model optimizer package MAGIX (Modeling and Analysis Generic
Interface for eXternal numerical codes), which helps to find the best
description of the data using a certain model \citep{moeller2013}. To
fit our data we used the XCLASS version 1.2.0 with CASA version 4.6.

\begin{figure}[htb]
\begin{center}
  \includegraphics[width=0.49\textwidth]{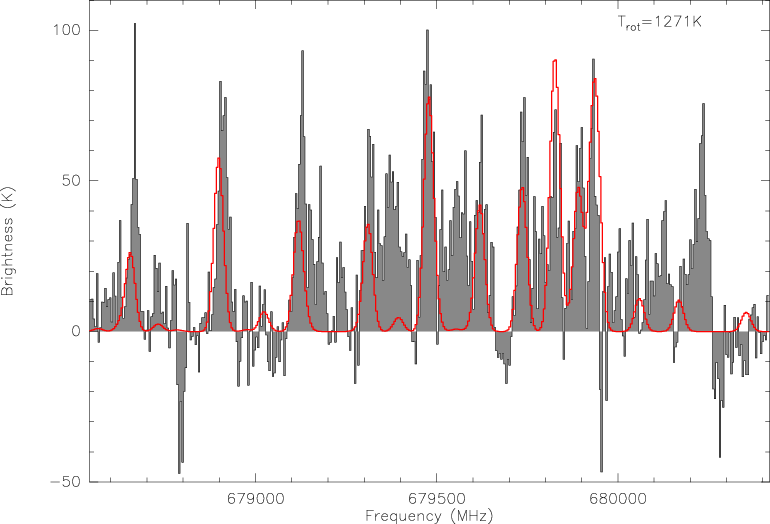}
\end{center}
\caption{Example CH$_3$CN$(37_k-36_k)$ spectrum and corresponding
  XCLASS two-component fit toward an arbitrary position of
  CH$_3$CN emission in the disk-like structure around source \#1. The
  corresponding temperature is marked at the top-right.}
\label{fit_example} 
\end{figure} 

To fit the data, we explored a broad parameter space and also
investigated whether single-component or double-component modeling
were more appropriate. The source structure with a central luminous
source implies that the temperature structure is more complicated, with
temperature gradients, and even shocks may contribute to the gas
temperature structure. However, increasing the fit complexity did not
improve the fit quality significantly. Temperature results for the
central area of interest varied by less than 10\% between the single-
and double-component fits. Nevertheless, to account at least partly
for the complexity, we resorted to a two-component fit.  Figure
\ref{fit_example} shows one example spectrum and the corresponding
XCLASS fit. While this fit itself does not appear great at first look,
the differences can largely be attributed to additional spectral
lines from other molecular species within the same spectral area.

In the following, we will continue our analysis with the warmer of the
two components because that resembles the innermost region
best. Figure \ref{temp} (middle panel) presents the derived hot
component temperature map for our central core/disk rotating
structure. The hole in the center of the map is caused by the
absorption against the strong continuum where no reasonable CH$_3$CN
fits are possible. We find very warm gas with temperatures largely
around 1000\,K with no steep gradient.  
%Colder gas components are difficult to identify with the
%high-excitation CH$_3$CN $K-$ ladder we are using here.

For comparison, in the bottom panel of Fig.~\ref{temp} we show also
the 438\,$\mu$m continuum map, this time converted to Planck
brightness temperatures $T_b$. Since we see all spectral lines in
absorption against the main peak, the central position may be
optically thick and thus could give an estimate of the dust
temperature. The highest value we find this way is 201\,K,
significantly below the gas temperatures estimated above from the
CH$_3$CN data. With the unknown exact optical depth, this dust
temperature should be considered a lower limit to the actual dust
temperature. Since the optical depth decreases going further out, this
assumption breaks down soon after leaving the main peak position. This
can also be tested by fitting profiles to the dust continuum map. We
explored this by fitting brightness temperatures to the west and east
from the center ($\Delta \rm{Offset} \propto T_b^{\alpha}$), and we
derived profile values $\alpha$ of $-1.7$ and $-1.6$,
respectively. This is far too steep for any reasonable disk or core
profile (usually $\alpha \sim -0.4$, e.g., \citealt{kenyon1987})) and
hence shows that the dust continuum map cannot be optically thick away
from the central position.

\section{Discussion}

\subsection{Fragmentation}

The number of submm continuum sources found toward G351.77-0.54 in
Figure \ref{cont} appears to be low within that field of
view. However, quantifying this more closely, we find 4 sources (\#1
through \#4) within a projected separation between \#3 and \#4 of
$\sim 7''$, corresponding to 15400(7000)\,AU at proposed distances of
2.2(1.0)\,kpc. Since the continuum optical depth is very high (see
section \ref{submm_cont}), the extended structure of source \#1 may
hide potential additional substructure. Independent of that, the
overall continuum structure resembles a proto-Trapezium-like system
(e.g., \citealt{ambartsumian1955,preibisch1999}). Assuming a spherical
distribution, this corresponds to source densities of $\sim 1.8\times
10^4$\,pc$^{-3}(2.0\times 10^5$\,pc$^{-3}$) at 2.2(1)\,kpc
distance. These densities compare or are even higher than source
densities found in typical embedded clusters (protostellar densities
typically around $10^4$\,pc$^{-3}$, \citealt{lada2003}). The
separations also compare well to the Orion Trapezium system (e.g.,
\citealt{muench2008}). Furthermore, similar high densities have also
been found in a few other high-mass star-forming regions, e.g., W3IRS5
\citep{rodon2008} or G29.96 \citep{beuther2007d}. Although our mass
sensitivity is very low $\leq 0.004$\,M$_{\odot}$, the comparably high
column density sensitivity of $5\times 10^{23}$\,cm$^{-2}$ (section
\ref{submm_cont}) indicates that some lower column density sources may
be hidden below our sensitivity estimates. Therefore, the above
estimated source densities should be considered as lower limits to the
actual densities in the region.

\subsection{(Disk) rotation?}
\label{discussion_disk}

As described in section \ref{dense} and shown in Figs.~\ref{mom1} and
\ref{pv} we identify clear velocity gradients perpendicular to the
main CO(6-5) outflow emanating from source \#1. However, the
position-velocity diagram also shows non-Keplerian contributions in
the two opposing quadrants of the pv-diagrams in Fig.~\ref{pv}. While
inclination effects can partly account for this (e.g.,
\citealt{seifried2016}), models of embedded disks within infalling and
rotating structures naturally result in such pv-diagrams (e.g.,
\citealt{ohashi1997,sakai2016,oya2016}).

While ideally one would like to estimate the mass of the central
object by fitting a Keplerian curve to the outer structure in the
pv-diagram (Fig.~\ref{pv}), this seems hardly feasible for G351.77. In
comparison to the fiducial 10(4.5)\,M$_{\odot}$ Keplerian curve (at
2.2(1.0)\,kpc presented in Fig.~\ref{pv}, we also show the
corresponding Keplerian curves for half and double the central
masses. Although one could argue that for example the $k=8, 9$
transitions may be best represented by the 10(4.5)\,M$_{\odot}$ curve,
this is less obvious in some of the other transitions (e.g., $k=5,
6$). Hence, determining the exact mass of the central object
unfortunately remains difficult.

\citet{leurini2011} note that there also seems to be CH$_3$CN emission
associated with the outflow, and if this is the case, the
interpretation becomes more complicated. We confirm this result in
that we see blue-shifted absorption along the line of sight as well as
a smaller scale red-shifted CO(6--5) emission feature in the
east-northeastern direction (Fig.~\ref{co}). It is interesting to note
that this is the case even for spectral lines with high excitation
temperatures (Table \ref{lines}). Furthermore, also outflows can
produce emission features in the other quadrants of the pv-diagrams
(e.g., \citealt{arce2007,li2013}).

Hence, although we find a clear velocity gradient perpendicular to an
outflow that is likely stemming to a significant fraction from an
embedded disk-like structure, there are several additional kinematic
signatures attributable to an infalling envelope as well as outflow
emission that complicate the overall interpretation.

\subsection{Gas and dust temperatures}
\label{temp_disc}

Why does the gas temperature derived from the CH$_3$CN$(37_k-36_k)$
$k-$ ladder deviate so much from the brightness temperature based on
the dust continuum emission? To first order, at the high densities
present in this region, one would expect that gas and dust should be
well coupled. For CH$_3$CN, \citet{feng2015} have shown that not
considering optical depth effects may cause significant overestimates
of estimated temperatures. However, since XCLASS takes also into
account the CH$_3^{13}$CN$(37_k-36_k)$ isotopologue emission, this
does not seem to be the case here. Therefore, the data indicate that
the CH$_3$CN and dust continuum emission both may trace different
physical entities within the region. While shocks can partially be
responsible for differences in the temperature structure, the gas and
dust may also trace different components of the expected underlying
massive disk.  Considering that both emission structures are dominated
by a disk-like structure around the central object and less affected
by the envelope because that is filtered out by the extended array
used here, the dust continuum emission may trace a colder disk
mid-plane whereas the CH$_3$CN emission may stem more from the disk
surface. While this scenario has been discussed for low-mass disks
(e.g., \citealt{henning2013,dutrey2014}), it is less well constrained
for disks around higher-mass young stellar objects. However, from a
physical point of view, one would expect similar
behavior. Investigating the best-fit model of the high-mass disk
source AFGL4176 by \citet{johnston2015} in more detail, such a
temperature structure is found qualitatively as well. We now may see
first observational evidence of that here in G351.77-0.54. In addition
to this, since the CH$_3$CN emission may be partly influenced by the
outflow, associated shocks may also contribute to the high
temperatures.

\subsection{Toomre stability of rotating structure}

Using the temperature information derived from the CH$_3$CN data, we
can also address the stability of the central rotating structure
around the main source \#1. Originally derived by \citet{toomre1964}
to estimate the gravitational stability of a disk of stars, it is used
since then in various environments for a stability analysis (e.g,
\citealt{binney2008}). In this framework, the Toomre $Q$ parameter
refers to the stability of a differentially rotating, gaseous disk
where gravity acts against gas pressure and differential rotation. The
Toomre $Q$ parameter is defined as

$$ Q= \frac{c_s \kappa}{\pi G \Sigma}$$

with the sound speed $c_s$, the epicyclic frequency $\kappa$ that
corresponds to the rotational velocity $\Omega$ in the case of a
Keplerian disk, and the surface density $\Sigma$. For thin disks,
axisymmetric instabilities can occur for $Q<1$. Broadening the
framework to disks with finite scale-height, the disks can stay stable
to slightly lower values because the thickness reduces the
gravity in the midplane where typically the structure formation takes
place. Studies with different disk density profiles found
critical $Q$ values below which axisymmetric instabilities occur
between $\sim$0.6 and $\sim$0.7 (e.g.,
\citealt{kim2007,behrendt2015}). In contrast to this, non-axisymmetric
instabilities like spirals or bars can also occur at slightly higher
$Q$ with $Q<2$ (e.g., \citealt{binney2008}). Taking these differences
into account, for $Q$ significantly higher than 2, the disks should
stay stable whereas at lower $Q$ non-axisymmetric and axisymmetric
instabilities can occur.

In the following, we estimate the Toomre $Q$ parameter for the whole
rotating structure around the main source \#1 calculating the sound
speed $c_s=\sqrt{kT/(2.3m_H)}$ with the temperature $T$
(Fig.~\ref{temp}) and the mass of the hydrogen nucleus $m_H$, and
using the Keplerian velocity $\Omega$ around a 10\,M$_{\odot}$
star. The assumption of Keplerian motion appears reasonable for the
inner disk considering the position-velocity diagram shown in
Fig.~\ref{pv}, and it may become invalid in the outer regions when the
disk mass becomes comparable to the mass of the central object and
gravitational instabilities can occur. Furthermore, we use our
438\,$\mu$m continuum map to estimate the column density $N_{{\rm
    H}_2}$ (or $\Sigma$) at each pixel with the assumptions outlined
in section \ref{submm_cont}.

\begin{figure}[htb]
\begin{center}
  \includegraphics[width=0.43\textwidth]{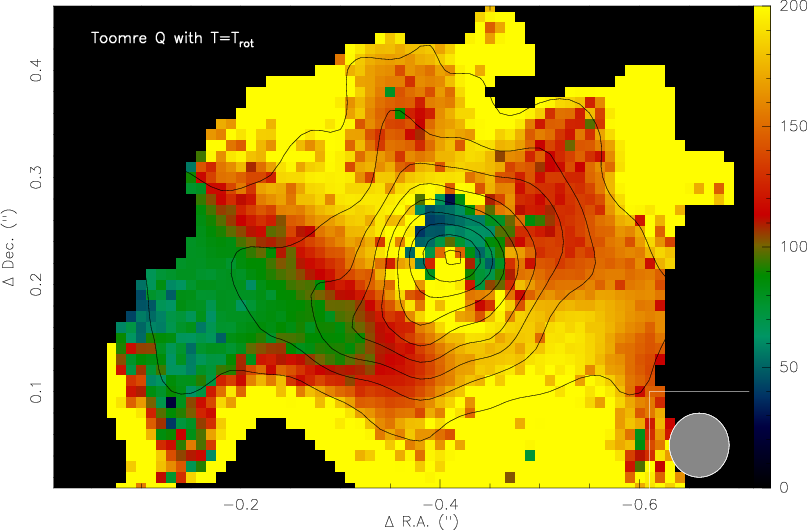}
  \includegraphics[width=0.43\textwidth]{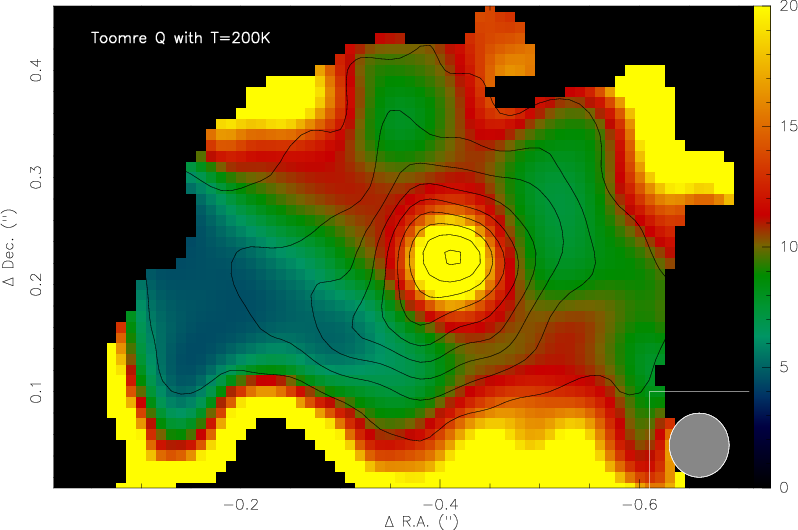}
  \includegraphics[width=0.43\textwidth]{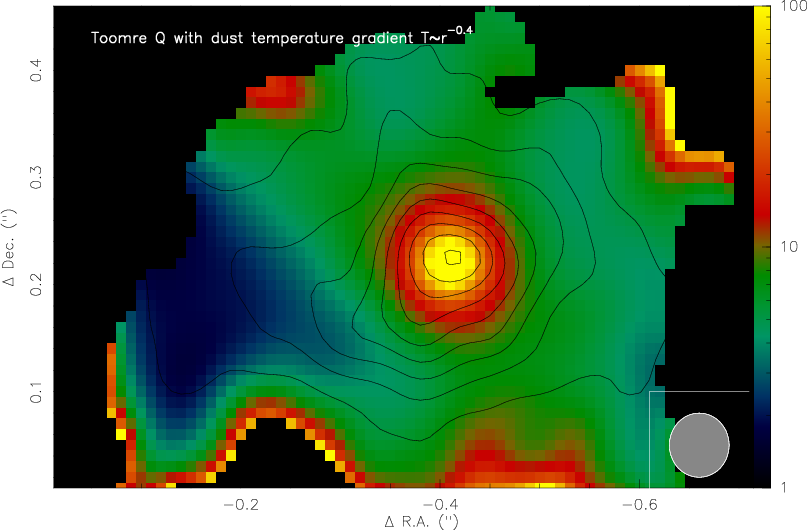}
\end{center}
\caption{The color-scales show in all three panels the Toomre Q map
  (see text for details) of the disk seen toward source \#1. While the
  top panel uses the rotational temperature derived from CH$_3$CN
  (Fig.~\ref{temp}) at each pixel to determine the sound speed and
  column density, the middle panel uses a uniform lower temperature of
  200\,K for these estimates. The lower panel uses a dust temperature
  distribution of $T\propto r^{-0.4}$ for the calculations (see
  section \ref{temp_disc}). The black contours show in each panel the
  438\,$\mu$m continuum emission in $4\sigma$ steps of
  28\,mJy\,beam$^{-1}$. Toomre Q values outside the
  CH$_3$CN$(37_2-36_2)$ 3$\sigma$ level (Fig.~\ref{temp}) are
  blanked. The beam is shown at the bottom-right of each panel.}
\label{toomre_plot} 
\end{figure}

Regarding the temperature for the sound speed and the column density
map, we present three different approaches: First, we use the CH$_3$CN
derived rotation temperature (Fig.~\ref{temp}) as a temperature proxy.
Second, we use a uniform lower temperature of 200\,K as temperature
proxy that may resemble the dust temperature better (sections
\ref{submm_cont} and \ref{temp_disc}). And third, we use a dust
temperature distribution with an approximate dust sublimation
temperature of 1400\,K at 10\,AU and then a temperature distribution
with radius $r$ as $T\propto r^{-\alpha}$. While $\alpha=0.5$
corresponds to the classical disk solution (e.g.,
\citealt{kenyon1987}), embedded disks can have lower $\alpha$ values
(e.g., the best-fit disk temperature profile for AFGL4176 has $\alpha
\sim 0.4$, \citealt{johnston2015}, Johnston et al.~in prep.), thus we
assume a value of $\alpha = 0.4$ for our analysis. Figure
\ref{toomre_plot} presents the corresponding Toomre $Q$ maps in the
three panels, with the third panel showing $Q$ for $T\propto
r^{-0.4}$. Since the temperature difference between these approaches
is large ($>1000$\,K and below 100\,K in the outskirts), we can
consider these Toomre $Q$ maps as the bracketing values for the real
physical Toomre $Q$ values in this region.

There is a large difference between the maps, with $Q$ values largely
$>$100 for $T=T_{\rm{rot}}$, between 5 and 20 for $T=200$\,K, and
between $\sim 1.5$ and several 100 for the temperature distribution
that varies as function of radius. However, even the lowest values in
the map derived with the temperature distribution are greater than
$\sim$1.5, hence in the axisymmetric stable regime. We note that with
a steeper temperature distribution of $\alpha=0.5$ the bluish parts in
the bottom panel of Fig.~\ref{toomre_plot} would shift below Toomre
$Q$ values of 1, but we consider such a steep distribution unlikely
considering the deeply embedded phase and the very high gas
temperatures measured in CH$_3$CN. Therefore, while we do see
sub-sources and fragments on larger scales (sources \#2 to \#4), the
central disk-like structure around the main source \#1 with an
approximate diameter of 1000(500)\,AU @2.2(1)\,kpc appears stable
against axisymmetric gravitational fragmentation. The main reason for
that is the high average temperature of the rotating structure. This
should at least be true for the resolved spatial scales. In addition
to this, for $Q$ values between 1 and 2 disks can be prone to
asymmetric instabilities such as spiral arms. Hence, it may still be
possible that locally even higher densities exist which then may allow
fragmentation on these smaller local scales.

\section{Conclusions}
\label{conclusion}

The Atacama Large Millimeter Array (ALMA) in band 9 (around 690\,GHz)
opens an entirely new window to study star formation and in particular
the formation of the highest mass stars. With correspondingly long
baselines we reach unprecedented spatial resolution of $\sim$0.06$''$,
allowing us to dissect the small-scale structure in high-mass
star-forming regions. Here, we present the first such study targeting
the high-mass star-forming region G351.77-0.54. 
%The luminosity of the
%region is $\sim 10^5$($\sim 1.7\times 10^4$)\,L$_{\odot}$ at
%2.2(1.0)\,kpc distance. 
With a linear resolution of 130(60)\,AU, we
resolve the fragmentation of the dense inner core in the 438\,$\mu$m
continuum emission as well as the kinematics and temperature structure
of the dense gas.

We resolve four submm continuum sources in the field of view which
result in approximate protostellar densities between $10^4$ and
$10^5$\,pc$^{-3}$ depending on the assumed distance of the
source. This should be considered as a lower limit since our column
density sensitivity is comparably high and we may miss lower column
density sources in the field. These high protostellar densities are
comparable to those found in other proto-Trapezia systems (e.g.,
\citealt{preibisch1999,rodon2008}).

The four spectral windows cover a multitude of spectral lines, here we
concentrate on the CH$_3$CN($37_k-36_k)$ $k-$ladder that covers
excitation levels above ground between 588 and 1300\,K, tracing
the warmest gas components in the region. Furthermore, we use the
CO(6--5) line wing emission to identify high-velocity gas emanating
from the main submm continuum source \#1 in the north-northwestern
direction. The kinematic analysis of the CH$_3$CN data identifies a
velocity gradient perpendicular to that outflow. While the
position-velocity gradients show hints of Keplerian rotation, there is
also an additional component that likely stems from the infalling
envelope as well as possible outflow contributions.

The continuum emission at these wavelengths is so strong with likely
high optical depth toward the peak positions that we identify
absorption in the dense gas tracer CH$_3$CN against the continuum
sources \#1 and \#2. While the absorption toward the main source \#1
is mainly blue-shifted and hence dominated by outflow motions, toward
the secondary unresolved source \#2, blue- and red-shifted absorption
is identified. These absorption features indicate high mass infall
rates on the order of $10^{-4}$ to $10^{-3}$\,M$_{\odot}$\,yr$^{-1}$
which can be considered as upper limit proxies for the actual mean
accretion rates onto the central source embedded in \#2.

Fitting simultaneously the CH$_3$CN($37_k-36_k)$ and CH$_3^{13}$CN
isotopologue $k-$ladders with XCLASS, we derive a rotational
temperature map for the main source \#1. Very high gas temperatures,
often in excess of 1000\,K are found. Hence, we are apparently seeing
part of the hottest molecular component observed so far in dense
high-mass star-forming regions. This is reasonable considering the
high excitation levels of the used lines. Assuming that at least the
central beam of the submm continuum emission is optically thick, one
can use its brightness temperature as a proxy of the dust
temperature. The highest continuum brightness temperature we find is
201\,K, indicating that we see a difference in gas and dust
temperature. We may witness here the difference between the colder
dust mid-plane and the hotter gaseous surface layer of a high-mass
accretion disk, similar to cases known for their low-mass
counterparts.

Analyzing the stability of the rotating structure via determining the
Toomre $Q$ parameter, we find the central disk-like structure to
exhibit always high $Q$ values, hence being on the resolvable scales
in the axisymmetrically stable regime. However, asymmetric
instabilities like spiral arms may still occur on even smaller so far
unresolved spatial scales.

\begin{acknowledgements} 
  This paper makes use of the following ALMA data:
  ADS/JAO.ALMA\#2013.1.00260.S. ALMA is a partnership of ESO
  (representing its member states), NSF (USA) and NINS (Japan),
  together with NRC (Canada) and NSC and ASIAA (Taiwan) and KASI
  (Republic of Korea), in cooperation with the Republic of Chile. The
  Joint ALMA Observatory is operated by ESO, AUI/NRAO and NAOJ. We
  like to thank Alvaro Sanchez-Monge and Thomas M\"oller for their
  continuous help with and improvements on the spectral line fitting
  program XCLASS. Furthermore, we thank the referee for a detailed
  report improving the clarity of the paper. HB acknowledges support
  from the European Research Council under the Horizon 2020 Framework
  Program via the ERC Consolidator Grant CSF-648505. RK acknowledges
  support from the German Research Foundation under the Emmy Noether
  Program via the grant no. KU 2849/3-1.

\end{acknowledgements}

\bibliography{/home/beuther/tex/bibliography}   
%\bibliography{/Users/henrikbeuther/tex/bibliography}
\bibliographystyle{aa}    % this does the style, aa.bst necessary

\end{document}